\documentclass[aps,pra,article,twocolumn,floatfix]{revtex4-1}

\usepackage[colorlinks]{hyperref}
\usepackage[utf8]{inputenc}
\usepackage{graphicx}
\usepackage{amsmath}
\usepackage{bm}
\usepackage[dvipsnames]{xcolor}
\usepackage[normalem]{ulem} %Cross-out text
\usepackage{subcaption}
\usepackage[format=plain,justification=centerlast]{caption}
\usepackage{lipsum} 
\usepackage{float}
\usepackage[absolute,overlay]{textpos}

\begin{document}

\title{Cooperative Cooling in a 1D Chain of Optically Bound Cold Atoms}
\author{Angel T. Gisbert}\email{angel.tarramera@unimi.it}
\affiliation{Dipartimento di Fisica "Aldo Pontremoli", Universit\`a degli Studi di Milano, Via Celoria 16, Milano I-20133, Italy}

\author{Nicola Piovella}\email{Nicola.Piovella@unimi.it}
\affiliation{Dipartimento di Fisica "Aldo Pontremoli", Universit\`a degli Studi di Milano, Via Celoria 16, Milano I-20133, Italy}

\author{Romain Bachelard}\email{bachelard.romain@gmail.com}
\affiliation{Universidade Federal de S\~{a}o Carlos, Rod. Washington Luis, km 235, S/n - Jardim Guanabara, S\~{a}o Carlos - SP, 13565-905, Brazil}

\date{\today}

\begin{abstract}
We discuss theoretically the optical binding of one-dimensional chains of cold atoms shone by a transverse pump, where particles self-organize to a distance close to an optical wavelength. As the number of particles is increased, the trapping potential increases logarithmically as the contributions from all atoms add up constructively. We identify a cooperative cooling mechanism, due to the mutual exchange of photons between atoms, which can beat the spontaneous emission for chains that are long enough. Surprisingly, the cooling is optimal very close to the resonance. This peculiar cooling mechanism thus gives new insights into the cooperative physics of low-dimensional cold atom systems.
\end{abstract}

\maketitle

\section{Introduction}

After the pioneering work by Ashkin on optical forces for micro-sized particles~\cite{Ashkin1970,Ashkin1986}, the manipulation of small objects using light beams has been applied successfully to a wide range of systems, from atoms~\cite{Chu1985,Hansch1975} to biological systems~\cite{Ashkin1987}. In this context, the role of the {\it interparticle} optical forces was soon noted~\cite{Thirunamachandran1980}. These forces can be either attractive or repulsive, depending on the specific distance between the scatterers, which suggests it can act as a mechanism for self-organization of the matter. Several years later, the self-organization of dielectric particles in suspension in a fluid was reported, with a pronounced preference for the particles to be separated by an integer number of optical wavelengths~\cite{Burns1989}. Coined Optical Binding (OB) at the time, it has since known various developments~\cite{Dholakia2010}.

OB can be realized using two main configurations: In the transverse one, the scatterers are spread in a plane orthogonal to the direction of propagation of the pump, and are submitted to a rather homogeneous phase and intensity profile~\cite{Yan2014}. In the longitudinal configuration, the pump propagates in the direction of the aligned scatterers~\cite{Guillon2006a}. In all cases, the coupling between the scatterers becomes increasing complex as their number increases, due to the propagation effects within the system. To circumvent these effects and generate longer bound chains, it has been proposed to resort to Bessel beams~\cite{Karasek2008,Karasek2009}.

OB relies on the trapping optical force overcoming the stochastic effect due to spontaneous emission. The trapping component is generally analyzed in terms of potentials, considering that each particle is trapped in a potential generated by the other scatterers. Finding stable configurations is then a self-consistent problem as moving a single scatterer affects the global stability of the system~\cite{Romero2008,DavilaRomero2010,Yan2014}. More generally, while a pair of scatterers tends to self-organize at a distance equal to an integer number of optical wavelengths, larger systems suffer from diffraction effects which alter this spacing, but also the system stability~\cite{Guillon2006}. Finally, despite the binding force scales poorly with decreasing scatterer size~\cite{Rodriguez2008}, OB has recently attracted a lot of attention for nanoparticles, as it appears as a potential mechanism for self-structuring at the nanoscale~\cite{Forbes2019}.

In this context, it was only discussed recently the possibility of binding optically cold atoms~\cite{Maximo2018}. Indeed, the binding force is comparatively stronger for particles of size comparable to the optical wavelengths~\cite{Rodriguez2008}, and the smallest objects such as atoms present unstable configurations as the heating due to the random recoil overcomes the binding potential~\cite{Gisbert2019}. Nevertheless, differently from dielectrics, cold atoms present an atomic resonance, which leads to an extra cooling mechanism for pairs of atoms in an OB configuration~\cite{Maximo2018,Gisbert2019}. Although this extra damping is not sufficient to reach stability for pairs of cold atoms without additional stabilization mechanism such as molasses, it represents a further step toward this goal.

In this theoretical work, we report on a cooperative cooling mechanism in a one-dimensional chain of cold atoms. The long-range nature of the light-mediated interaction manifests not only in the deepening of the OB potential, but also in the enhancement of the cooling mechanism for resonant scatterers. Differently from other cooling mechanisms, including the cooling of a pair ($N=2$) of optically bound atoms, the cooling for larger systems ($N\geq3$) is most efficient at or very close to the atomic resonance, and in the 1D chain under study, it grows logarithmically with the system size. This self-generated cooling makes stable OB possible for chains of a few dozens of cold atoms. Our result shows that cooperative effects in low-dimensional cold-atom systems may be particularly useful for self-organization processes.

\section{Optical binding in a chain of cold atoms}

\subsection{Modeling the atomic chain dynamics}

\begin{figure}[h!]
\centering
\includegraphics[width=0.6\columnwidth]{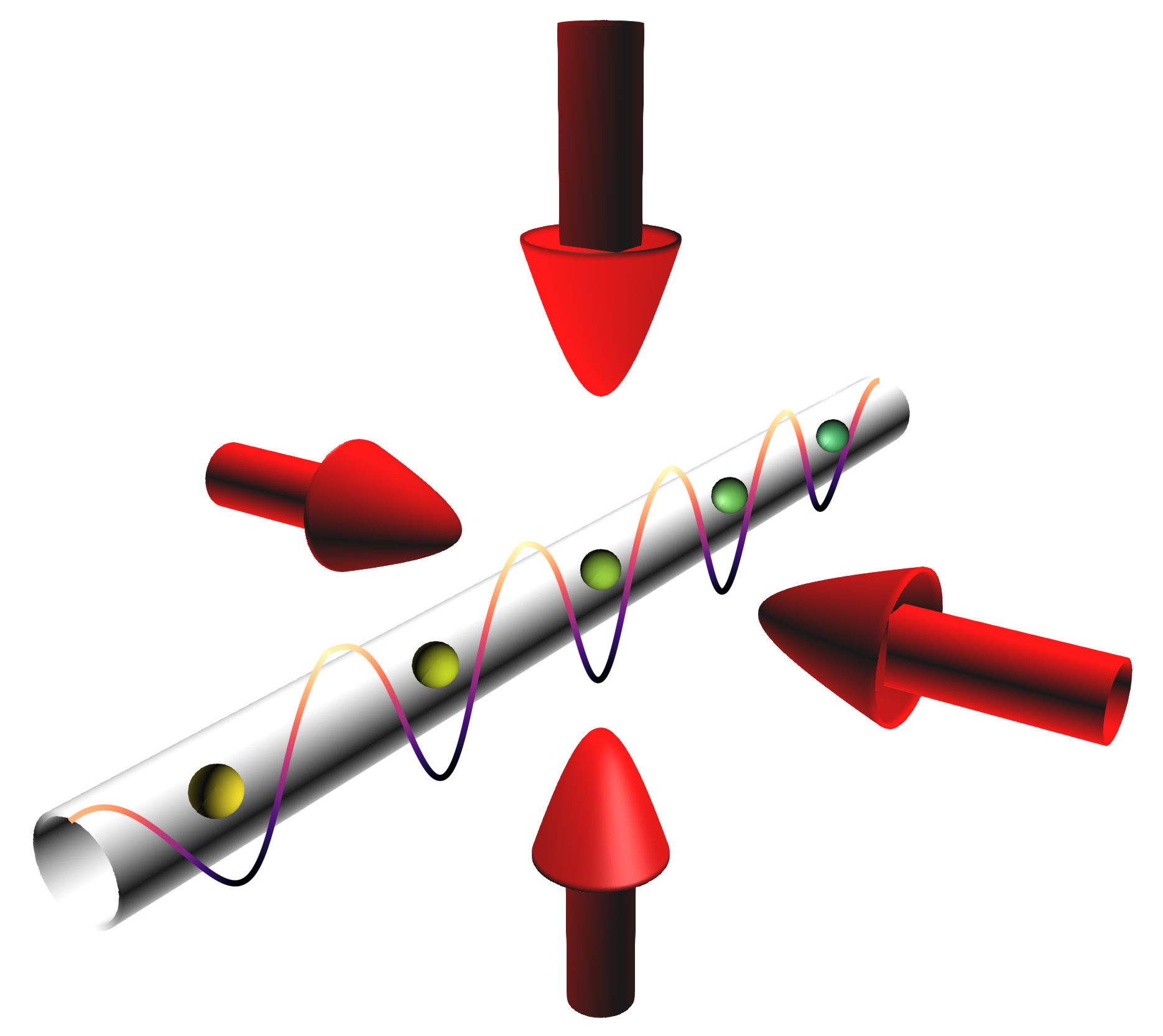}
\caption{Scheme of the one-dimensional cloud of atoms trapped by four laser beams, the self-organization as a chain occurring under the effect of mutual optical forces. The strongest coupling between neighbors is achieved when the mutual distance is close to the optical wavelength $\lambda$. The four laser beams drawn form a 2D optical lattice in the transverse directions and aim to emulate a one-dimensional system: they are ideally far from resonance. Differently, the near-resonant pump which generates the OB is also transverse, so it should be operated on a different transition.}
\label{fig:sketch}
\end{figure}

The dynamics of the optical binding involves monitoring the coupled evolution of both the vacuum modes and the atoms internal and external degrees of freedom. In order to reach an efficient description of the system, we focus on the atom dynamics, by tracing over the degrees of freedom of the light and studying the coupled dipole dynamics~\cite{Lehmberg1970,Purcell1973,Draine1988}. Considering we are dealing with two-level atoms, the dynamics of their dipoles, hereafter labeled $\beta_j$ and treated classically, is given by:
\begin{equation}
\frac{d\beta_j}{dt}=\left(i\Delta-\frac{\Gamma}{2}\right)\beta_j-i\Omega(\mathbf{r}_j)-\frac{\Gamma}{2}\sum_{l\neq j}G_{jl}\beta_l,
\label{eq:betaj}
\end{equation}
where $\Gamma$ is the linewidth of the atomic transition, $\Omega$ the Rabi frequency of the driving field, and $\Delta = \omega - \omega_a$ the detuning of the pump field from the atomic transition frequency $\omega_a$. 
We consider a setup of transverse one-dimensional OB, where the atoms are trapped in one dimension by a 2D optical lattice created by four plane-wave beams in the orthogonal plane (see Fig.\ref{fig:sketch}). Such a scheme allows to reduce the cold atom dynamics to one dimension and has been explored in various experiments~\cite{Gorlitz2001,Paredes2004,Glicenstein2020}.

 It corresponds to a pump with an homogeneous phase along the chain: $\Omega(\mathbf{r})=\Omega_0$. The light-mediated interaction between the dipole is given by the kernel $G_{jl}=\exp(ik|\mathbf{r}_j-\mathbf{r}_l|)/(ik|\mathbf{r}_j-\mathbf{r}_l|)$, where $\mathbf{r}_j$ refers to the position of the atom center of mass and $k=2\pi/\lambda$ the light wavenumber. This kernel can be seen as referring to scalar dipoles (scalar light approximation), or to vectorial dipoles oriented at a magic angle such that near-field terms cancel (i.e., a pump polarization which makes an angle $\theta=\arcsin{(1/\sqrt{3})}$ from the chain axis). In the linear-optics regime considered throughout this work, this dynamics can be obtained either from a quantum description of the light-matter interaction~\cite{Lehmberg1970} or from a representation of the atoms as classical oscillators~\cite{Svidzinsky2010, Cottier2018}. % This 'coupled dipole equations' has been studied extensively in the case of motionless atoms, as it captures coop effects REFS, but also the disorder-induced Anderson transition for light~\cite{Skipetrov2014}REFS.

Regarding the atom center-of-mass, their dynamics is driven only by the field from the other dipoles, since the trapping beams do not induce any force along the $z$-axis:
\begin{eqnarray}
m\frac{d^2\mathbf{r}_j}{dt^2} &=& -\hbar \left(\beta_j^*\nabla_{\mathbf{r}_j}\Omega_j+c.c.\right)
\nonumber
\\ &=& -\hbar\Gamma\sum_{l\neq j}\mathrm{Im}\left(\beta_j^*\beta_l\nabla_{\mathbf{r}_j}G_{jl}\right),
\label{eq:rj}
\end{eqnarray}
with $\Omega_j(\mathbf{r}_j)=\Omega_0-i(\Gamma/2)\sum_{l\neq j}G_{jl}\beta_l$ the effective Rabi frequency at position $\mathbf{r}_j$, $m$ the atom mass and $\hbar$ the reduced Planck constant. In this equation, stochastic effects associated to spontaneous emission have been eliminated -- see Sec.\ref{sec:fluct} for a more detailed discussion.

\subsection{Adiabatic approximation}

Systems of dielectrics previously considered for OB do not possess a resonance like atoms; it is equivalent to considering that the internal degrees of freedom, here the $\beta_j$s, are always at equilibrium. Performing such an adiabatic approximation, i.e., considering that the dipole relaxation time $\Gamma^{-1}$ is negligible compared to the time needed for an atom center of mass to perform an oscillation in the binding potential, corresponds to taking the left-hand term in Eq.~\eqref{eq:betaj} equal to zero. This allows to rewrite the dipole as $\beta_j=\alpha \Omega_j$, with $\alpha=1/(\Delta+i\Gamma/2)$ the normalized atom polarizability. Defining $\Omega_j=|\Omega_j|e^{i\varphi_j}$, the centers of mass dynamics in turn rewrites as:
\begin{eqnarray}
m\frac{d^2\mathbf{r}_j}{dt^2} &=&-\hbar\left(\alpha^*\Omega_j^*\nabla_{\mathbf{r}_j}\Omega_j+c.c.\right)\label{eq:rjadiab}\\ 
&=& \frac{\hbar\Gamma}{\Delta^2+\Gamma^2/4}|\Omega_j|^2\nabla_{\mathbf{r}_j}\varphi_j-\frac{\hbar\Delta}{\Delta^2+\Gamma^2/4}\nabla_{\mathbf{r}_j}|\Omega_j|^2,\nonumber
\end{eqnarray}
where the first right-hand term corresponds to the radiation pressure force, and the second to the dipolar force. %Thus, in the adiabatic approximation the dynamics of the centers of mass is conservative, with the (optical) potential energy given by . In particular, it means the ... energy conserved

\begin{figure}[ht!]
    \includegraphics[width=\linewidth]{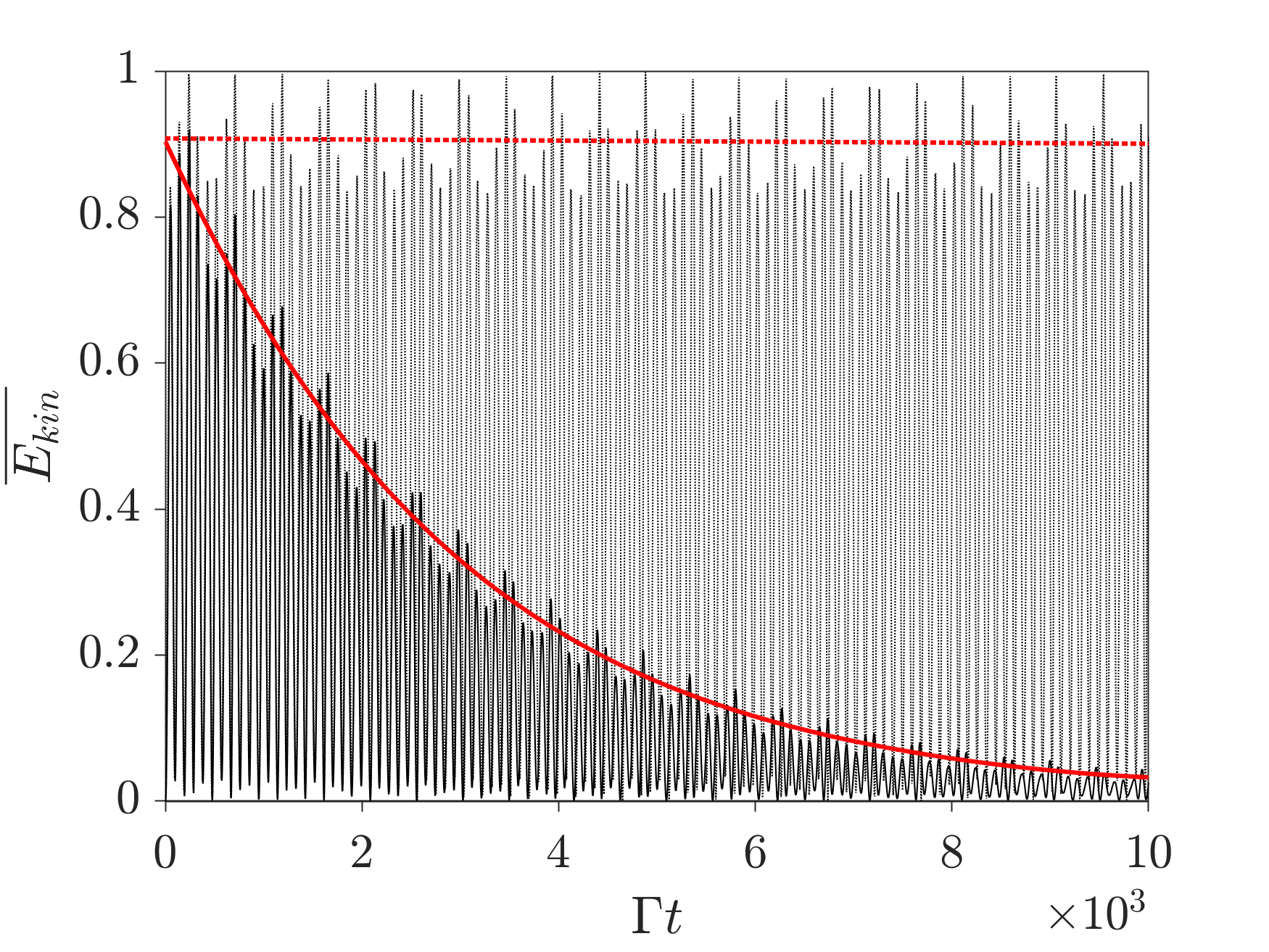}
    \caption{Evolution of the normalized kinetic energy for a chain of $N=15$ atoms with an initial inter-particle distance $\lambda$, and pumped with a laser detuned by $\Delta\approx-0.2\Gamma$. The darker black curve refers to the evolution with the full dipole dynamics, Eqs.(\ref{eq:betaj}) and (\ref{eq:rj}), the lighter curve was obtained from the adiabatic dynamics (cancelling the left-hand term in Eq.(\ref{eq:betaj})), and the red lines refer to the envelope obtained by averaging over a short time. The two kinetic energies have been normalized by the maximum of both curves.}
    \label{fig:Ekin}
\end{figure}

Despite we are dealing with an open system, for a pair of atoms ($N=2$) and after a short transient, the dipoles synchronize and the adiabatic approximation can be mapped to a conservative dynamics, derived from a potential energy~\cite{Gisbert2019}. This conveniently allows to monitor the evolution of the effective energy of the system. Differently, for a many-atom chain ($N\geq3$) the absence of synchronization translates into different dipole amplitudes, which in turn prevents defining a potential energy for the system: The mutual radiation pressure terms  (i.e., the phase gradient term in Eq.\eqref{eq:rjadiab}) cannot be expressed as deriving from a potential. Nevertheless, simulations of the kinetic energy with and without the adiabatic approximation show that performing this approximation leads to a conservative dynamics: On the time scales over which the system otherwise cools/heats, no significant long-term evolution of the kinetic energy is observed for the adiabatic dynamics, see Fig.~\ref{fig:Ekin}.

\subsection{Local potential at equilibrium}
In this work, the system is prepared out of equilibrium as follows: A chain of atoms separated by $\lambda$ is generated; the atom positions corresponding to the minima of the optical potential are then obtained by letting the system relax in presence of an artificial friction force $-\xi(d\mathbf{r}_j/dt)$ applied to all atoms. These minima correspond to a separation of the atoms slightly different from $\lambda$, and must be found as a self-consistent problem where all scatterers mutually interact~\cite{DavilaRomero2010}: The friction allows to reach the equilibrium in an efficient way. We have checked that the equilibrium positions are not affected by the value of $\xi$, which we have set to $0.02m\Gamma$ throughout this work. Then, the two atoms at the extremity of the chain are shifted away by $3\%$ of the distance to their nearest neighbour. The dynamics is then initiated with the atoms in these positions, without any initial velocity. 

In this review, we have simulated Eqs.(\ref{eq:betaj}-\ref{eq:rj}) using $\Omega_0=0.1\Gamma$ and $\omega_{\mathrm{rec}}=0.045\Gamma$, where $\omega_{\mathrm{rec}}=\hbar k^2/2m$ is the recoil frequency. This value of $\omega_{\mathrm{rec}}$ is low enough to neglect the shift induced by the scattering on the light frequency, yet large enough to observe the cooling over dozens of oscillations (lower ratios $\omega_{\mathrm{rec}}$ lead to larger time scales for the cooling~Ref.\cite{Gisbert2019}).

\begin{figure}[ht!]
    \centering
    \begin{subfigure}[b]{0.48\textwidth}
        \includegraphics[width=\columnwidth]{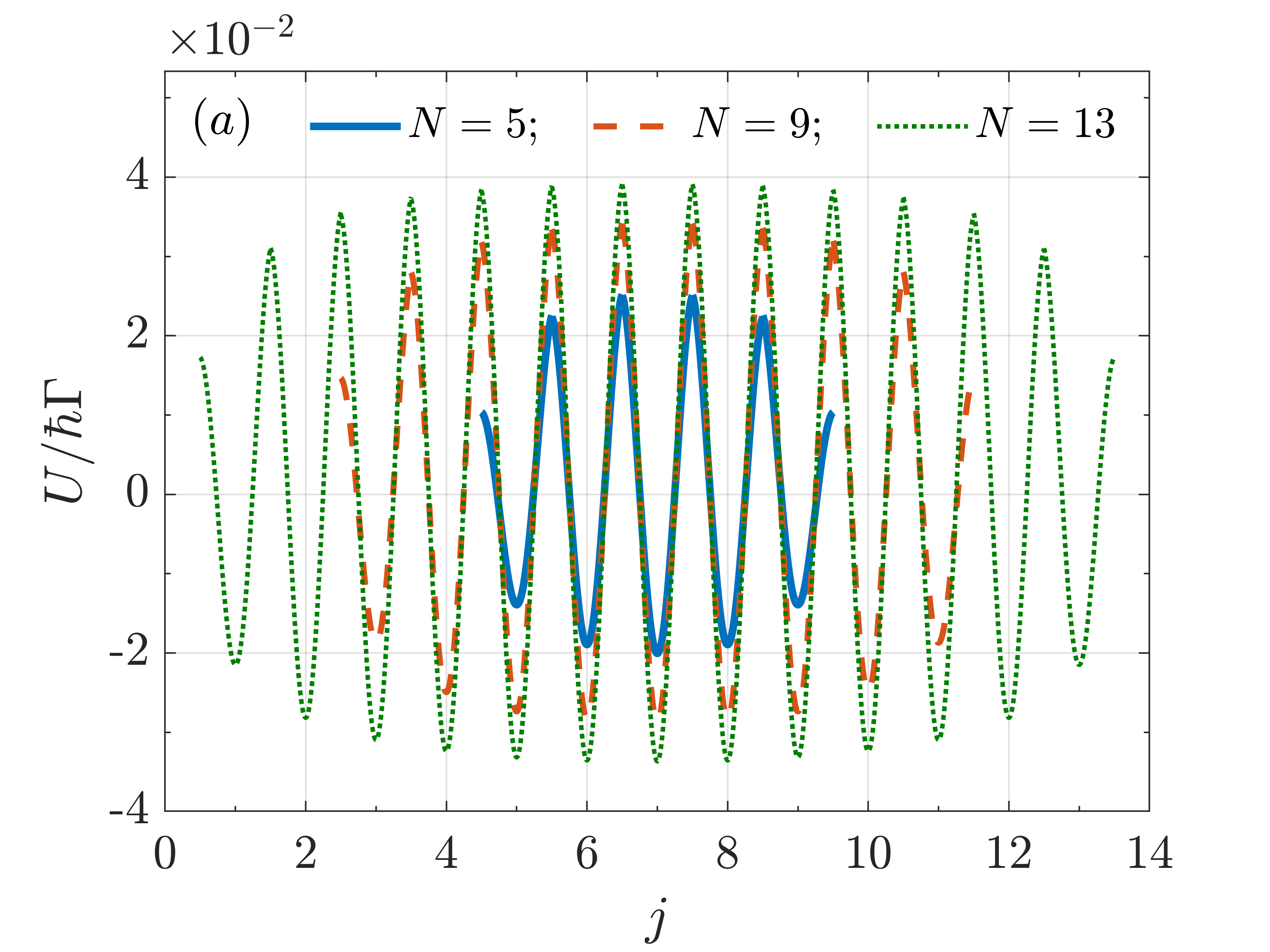}
    \end{subfigure}
    \begin{subfigure}[b]{0.48\textwidth}
        \includegraphics[width=\columnwidth]{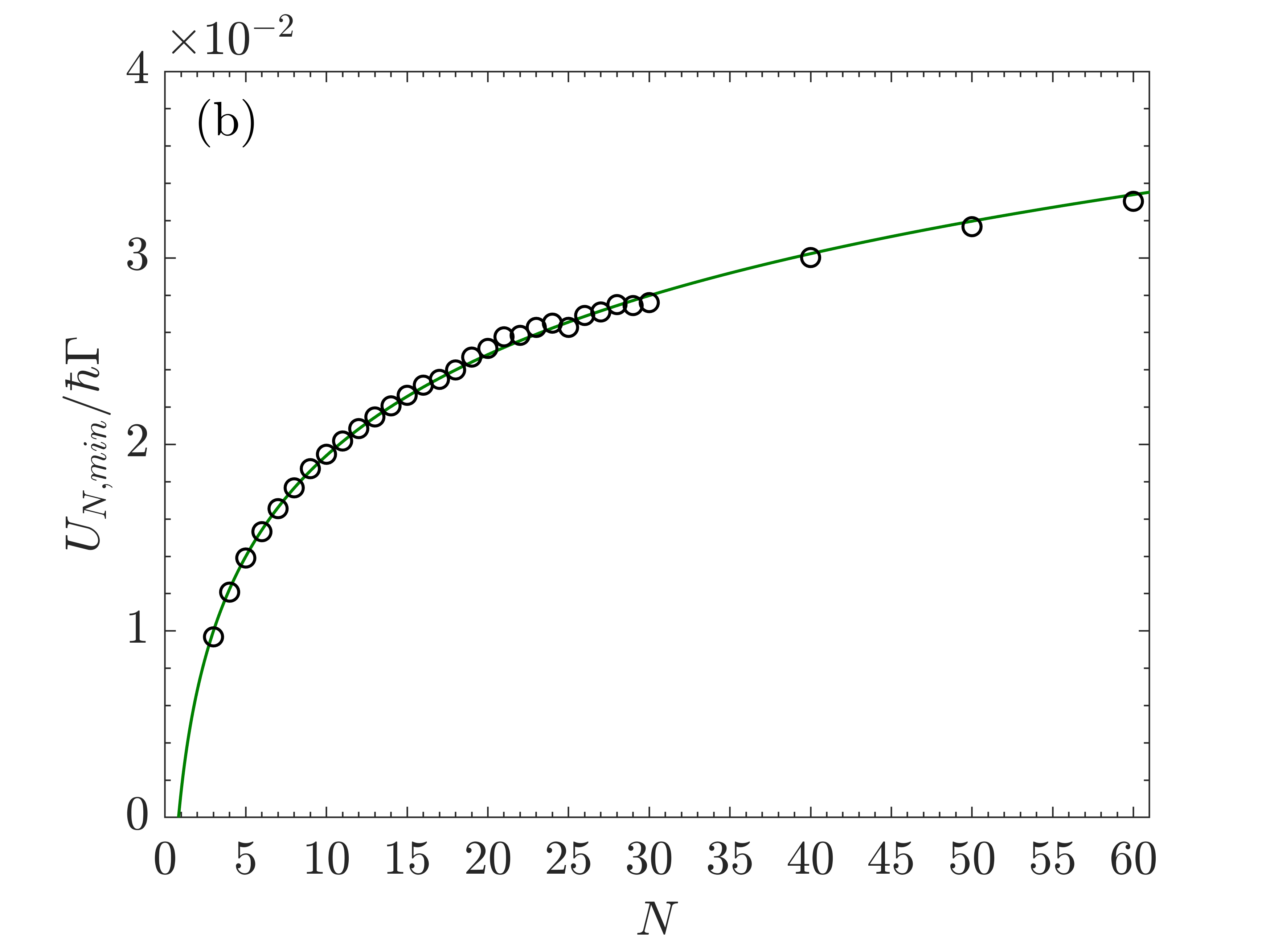}
    \end{subfigure}
    \caption{(a) Potential energy landscape for chains of $N=5,\ 9$ and $13$ atoms at equilibrium, for a normalized detuning $\Delta/\Gamma= -0.06,\ -0.13$ and $-0.18$, respectively. The potential is computed using Eq.\eqref{eq:Uj}, considering all atoms apart from the closest one, as it generates a local singularity. (b) Optical potential (in absolute value) for the edge atoms of a chain of length $N$, as a function of $N$, and for a detuning that optimizes the cooling. The green line corresponds to a logarithmic fit.}
    \label{fig:potential}
\end{figure}

The OB potential for each atom strongly depends on the system size. Indeed, due to the long-range nature of the dipole-dipole interaction, all atoms contribute to the instantaneous potential $U_j$ for atom $j$, which is deduced from Eq.~\eqref{eq:rj} as:
\begin{equation}
U_j =\hbar\Gamma\sum_{l\neq j}\mathrm{Im}\left(G_{jl}\beta_j^*\beta_l\right).
\label{eq:Uj}
\end{equation}
Let us discuss the potential generated by atoms once they have reached the minimum of the optical potential (since, in practice, the OB potential is a dynamical quantity). As can be observed in Fig.~\ref{fig:potential}(a), the optical potential for each atom becomes deeper as the chain size increases. This effect can be understood from the $1/r$ decay of the electric field. If, for simplicity, we assume that each atomic dipole is mainly driven by the laser field, $\beta_j=\Omega_0/(\Delta+i\Gamma/2)$, then the optical potential reads
\begin{equation}
    U_j=\hbar\Gamma\frac{\Omega_0^2}{\Delta^2+\Gamma^2/4}\sum_{l\neq j}\mathrm{Im}(G_{jl}).
\end{equation}
Assuming that all the atoms are separated by $\lambda$ (i.e., $\mathbf{r}_j=j\lambda\hat{z}$), the potential simplifies into
\begin{equation}
    U_j= U_0\sum_{l\neq j}\frac{1}{|l-j|}=C_j U_0,\label{eq:Usimp}
\end{equation}
with $U_0=-\hbar(\Gamma/2\pi)\Omega_0^2/(\Delta^2+\Gamma^2/4)$ the potential minimum for a pair of atoms, and $C_j=\sum_{l\neq j}1/|l-j|$ the cooperativity parameter for atom $j$. Thus, in a long chain, an atom in the middle of the center is submitted to a potential that is the coherent sum of the contributions of all atoms, the overall potential scaling as $$\sim U_0\sum_{j=-N/2,j\neq 0}^{N/2}1/|j|\sim 2U_0\ln (N/2).$$ An atom at the chain border is submitted to a smaller potential, $U_1=U_N\sim U_0\ln N$. This explains the scaling observed in Fig.~\ref{fig:potential}(a), which clearly favors larger chains in terms of stability.

In Fig.~\ref{fig:potential}(b), the logarithmic growth of the potential can be observed, for chains up to $N=60$ atoms. A fit of the numerically computed potential minimum  for the atoms at the extremes of the chain gives the following approximated expression:
\begin{equation}
U_{min}\approx -0.8\hbar\Gamma\left(\frac{\Omega_0}{\Gamma}\right)^2\ln N.\label{Umin}
\end{equation}
A slight decrease in the potential depth is observed for the largest system sizes, which can be explained from the finite optical thickness which separates remote atoms in long chains. Indeed, the exchange of photons between two remote atoms is screened by the in-between atoms, which modify both the amplitude and phase of the wave. As a result, the coherent sum \eqref{eq:Usimp} is no longer valid. Longer chains obviously suffer stronger screening effects, which represents a limit to the length of optically bound chains. In order to overcome such effect and bind efficiently long chains of scatterers, it has, for example, been proposed to shrink the coherence of the incident field to reduce the number of coherently interacting dipoles, or to spatially modulate the phase of the incident field~\cite{Guillon2006}.

\section{Cooperative cooling}

As mentioned earlier, an important difference of cold atoms as compared to dielectrics spheres is the presence of a resonance: It makes the atomic dipoles have a finite-time response to the local electric field. As a consequence, the system may either present a long-term cooling or heating trend~\cite{Maximo2018, Gisbert2019}, beyond the adiabatic dynamics described above. This is illustrated in Fig.~\ref{fig:Ekin}, where the evolution of the kinetic energy of the atomic chain presents a slow decay when the dipole dynamics is accounted for. The oscillations observed occur at a frequency provided by the trapping potential, which can be estimated from Eq.~\eqref{eq:Usimp}: 
\begin{equation}
    \omega_j^2=\frac{\omega_{\mathrm{rec}}\Gamma}{\pi}\frac{\Omega_0^2}{\Gamma^2+4\Delta^2}\sum_{l\neq j}\frac{1}{|l-j|}.
\end{equation}
where $\omega_{\mathrm{rec}}=\hbar k^2/2m$ is the recoil frequency. This frequency contains the cooperativity parameter $C_j=\sum_{l\neq j}1/|l-j|$, which scales as $\ln N$.

\begin{figure}[!hb]
    \begin{subfigure}[b]{0.225\textwidth}
    \includegraphics[width=\textwidth]{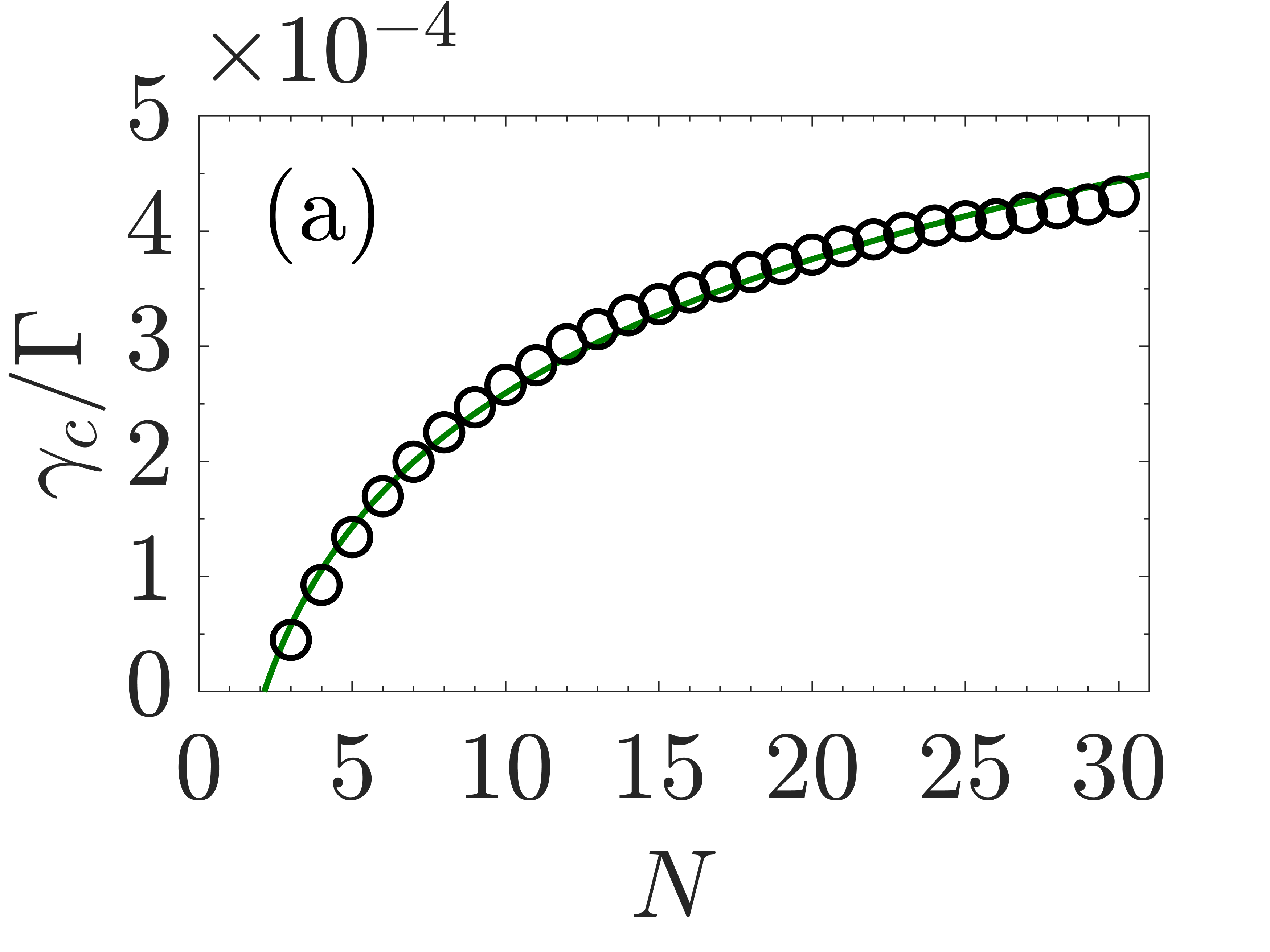}
    \label{fig:f2}
  \end{subfigure}
  \begin{subfigure}[b]{0.225\textwidth}
    \includegraphics[width=\textwidth]{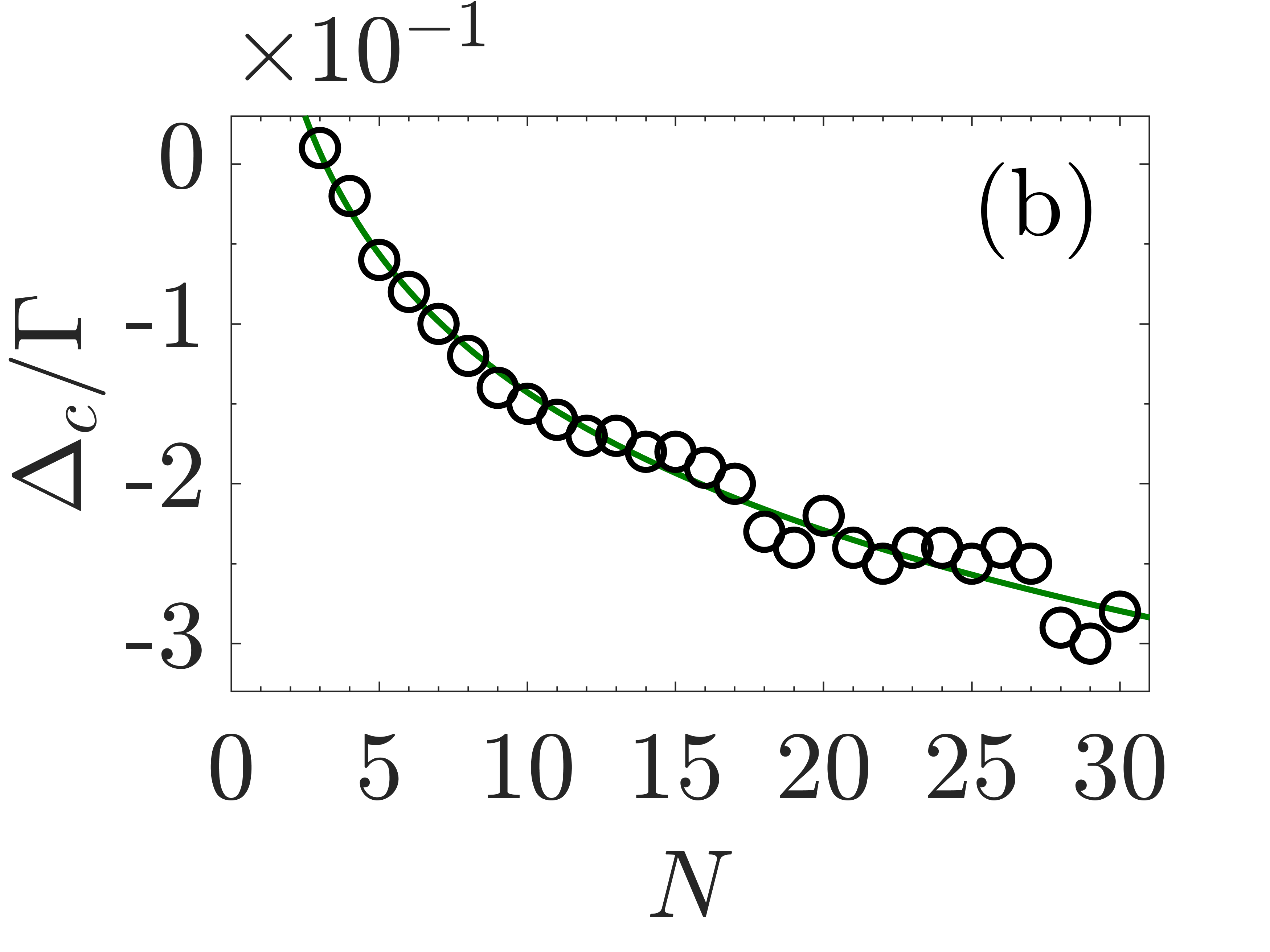}
    \label{fig:f1}
  \end{subfigure}
  \begin{subfigure}[b]{0.485\textwidth}
    \includegraphics[width=\textwidth]{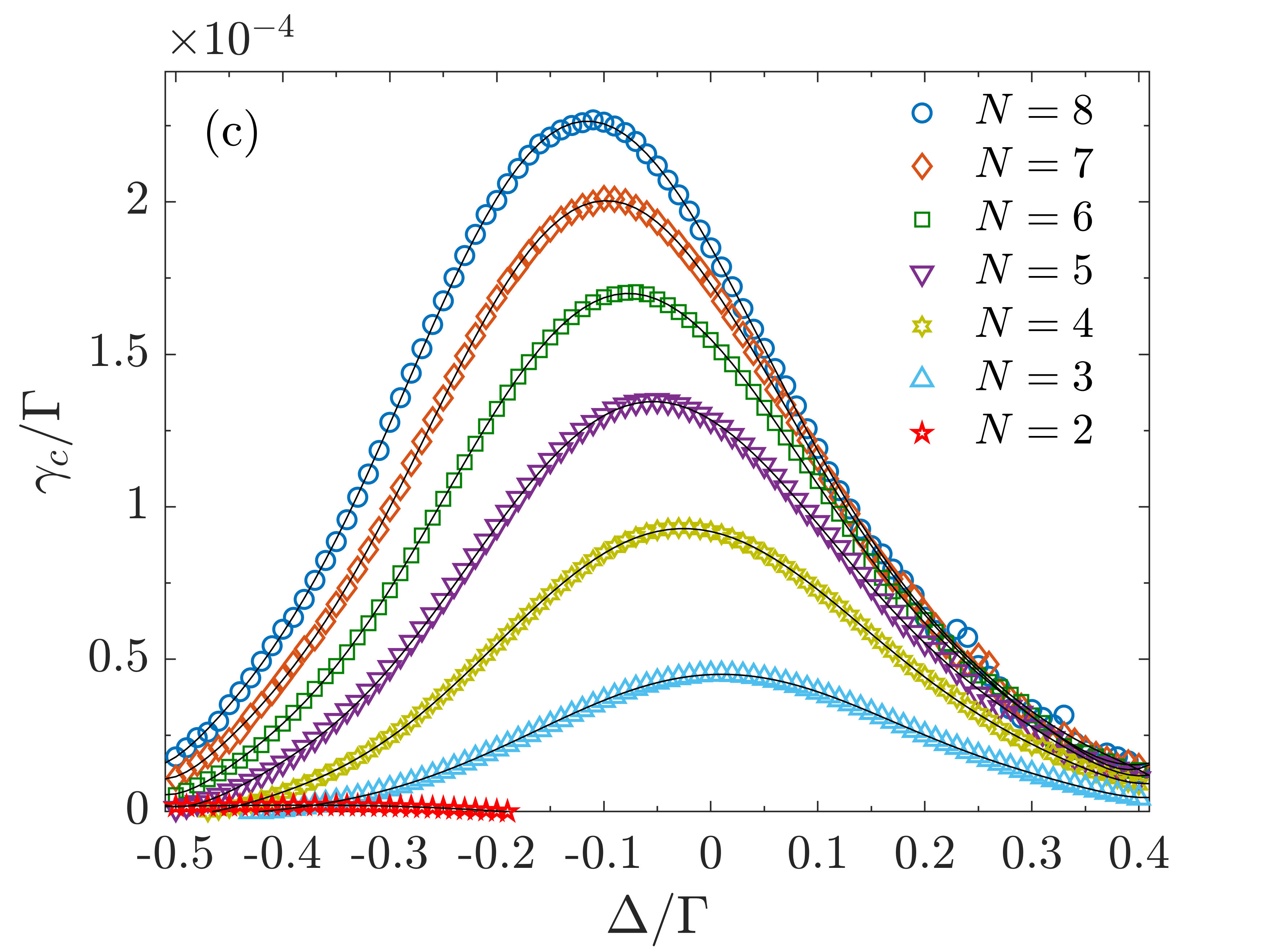}
    \label{fig:f3}
  \end{subfigure}
  \caption{(a) Maximum cooling rate $\gamma_c$ as a function of the particle number $N$.  (b) Detuning $\Delta_c$ of maximum cooling rate, as a function of  $N$. (c) Cooling rate $\gamma_c/\Gamma$ as a function of the normalized detuning $\Delta/\Gamma$ of the pump light and for different particle numbers.}
  \label{fig:gammac}
\end{figure}

The cooling observed in Fig.~\ref{fig:Ekin}, obtained from the envelope of the kinetic energy $\overline{E}$, is exponential in time, so we deduce a cooling rate $\gamma_c$ by an exponential fit. 
%We associate with it a cooling time $\tau_c=1/\gamma_c$, such that the kinetic energy reaches half of its initial value. 
The dependence of this rate on the particle number $N$ is presented in Fig.~\ref{fig:gammac}, for the detuning $\Delta_c$ that optimizes this rate (see discussion below).
However, we first remark that the cooling rate, $\gamma_c$, scales with $\ln N$ (see panel (a)), leading to an increased cooling rate for larger systems. This cooperative enhancement of the cooling, and the detuning $\Delta_c$ that optimizes this rate, are given by the following expressions, obtained by numerical fit:
\begin{eqnarray}
\gamma_c &\approx& \omega_{\mathrm{rec}}\left(\frac{\Omega_0}{\Gamma}\right)^2\left[0.4 \ln N-0.3\right],\label{gamma_c}
\\ \frac{\Delta_c}{\Gamma}&\approx& 0.14-0.12 \ln N,\label{Delta_c}
\end{eqnarray}
We remind that in the present setup there is no external damping force such as fluid friction for scatterers maintained in fluids~\cite{Burns1989}. Furthermore, in the case of transverse OB, the pump laser confine the particles only along the chain, and have no direct role on the dynamics along that direction: only interparticle optical forces contribute here. 

\begin{figure}[h!]
    \centering
    \includegraphics[width=\linewidth]{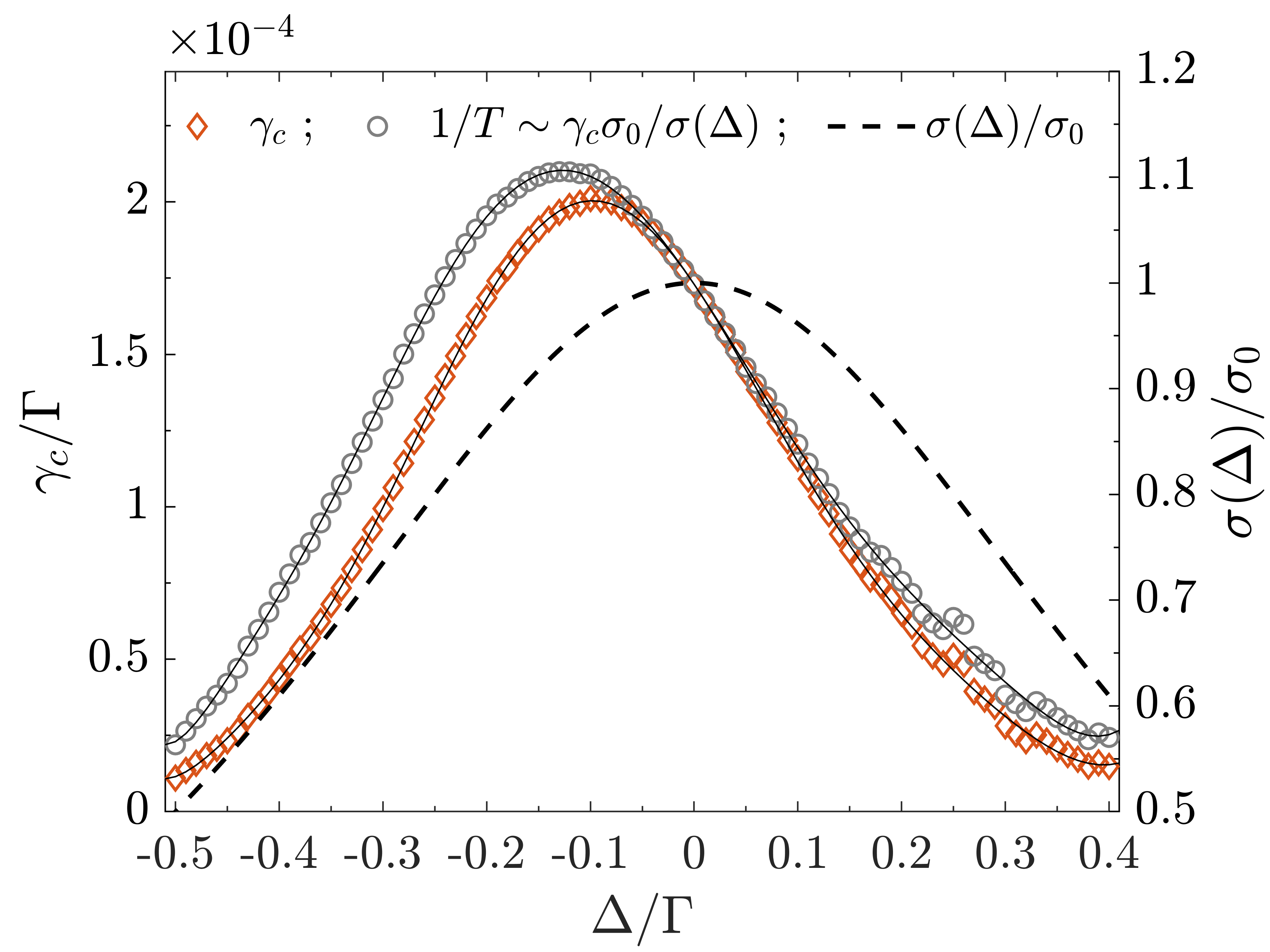}
    \caption{Cooling rate $\gamma_c/\Gamma$, scattering cross-section $\sigma$ and inverse temperature $1/T$ for a chain of $N=7$ atoms and as a function of the detuning of the pump light $\Delta$.}
    \label{fig:CRvsCS}
\end{figure}

In Fig.\ref{fig:gammac}(c) we observe a maximum cooling rate very close to resonance, nevertheless the steady-state is also determined by the spontaneous emission rate, which is also maximum at resonance. Indeed resonant light corresponds to a maximum scattering cross-section $\sigma(\Delta)=\sigma_0/(1+4\Delta^2/\Gamma^2)$, with $\sigma_0=4\pi/k^2$ the resonant scattering atom cross-section. The equilibrium temperature resulting from the cooling and the stochastic heating is obtained using a Langevin equation, and scales as $T\propto\sigma(\Delta)/\gamma_c(\Delta)$ (see Sec.\ref{sec:fluct}). As can be observed in Fig.~\ref{fig:CRvsCS}, despite the increased spontaneous emission, the equilibrium temperature is predicted to be lowest very close to resonance. The cooling thus appears to rely on the radiation pressure force rather than on the dipolar one. 

This makes the cooling mechanism for large ($N\geq3$) optically bound atomic chains quite different from other cooling mechanisms. In the case of an optically-bound pair of atoms~\cite{Maximo2018,Gisbert2019}, the scaling on the cooling rate was similar to the one obtained for Doppler cooling: Cooling is achieved for negative detuning, ideally for $\Delta\approx-\Gamma/2$, whereas positive detuning is associated with heating (see Fig.~\ref{fig:gammac}(c)). For $N\geq 3$ atoms in an OB configuration, the cooling not only appears most efficient very close to resonance, and even scales differently from the $N=2$ case: It reaches a maximum $\sim 20$ times higher for $N=3$ than the maximum reached for a pair of atoms. Only for larger numbers does the optimal pump frequency start to deviate from the atomic resonance ($\Delta_c\approx\Gamma/4$ for $N=20$, see panel (c) on Fig.~\ref{fig:gammac}). The present situation is at odds from the cooling by diffuse light reported for atoms trapped in a reflecting cylinder~\cite{Guillot2001,Cheng2009,LingXiao2010}, where the cooling was achieved in a fully disordered system, and was optimal off-resonance ($\Delta\approx -3\Gamma$). 

\begin{figure}[ht!]
\centering
\includegraphics[width=1\columnwidth]{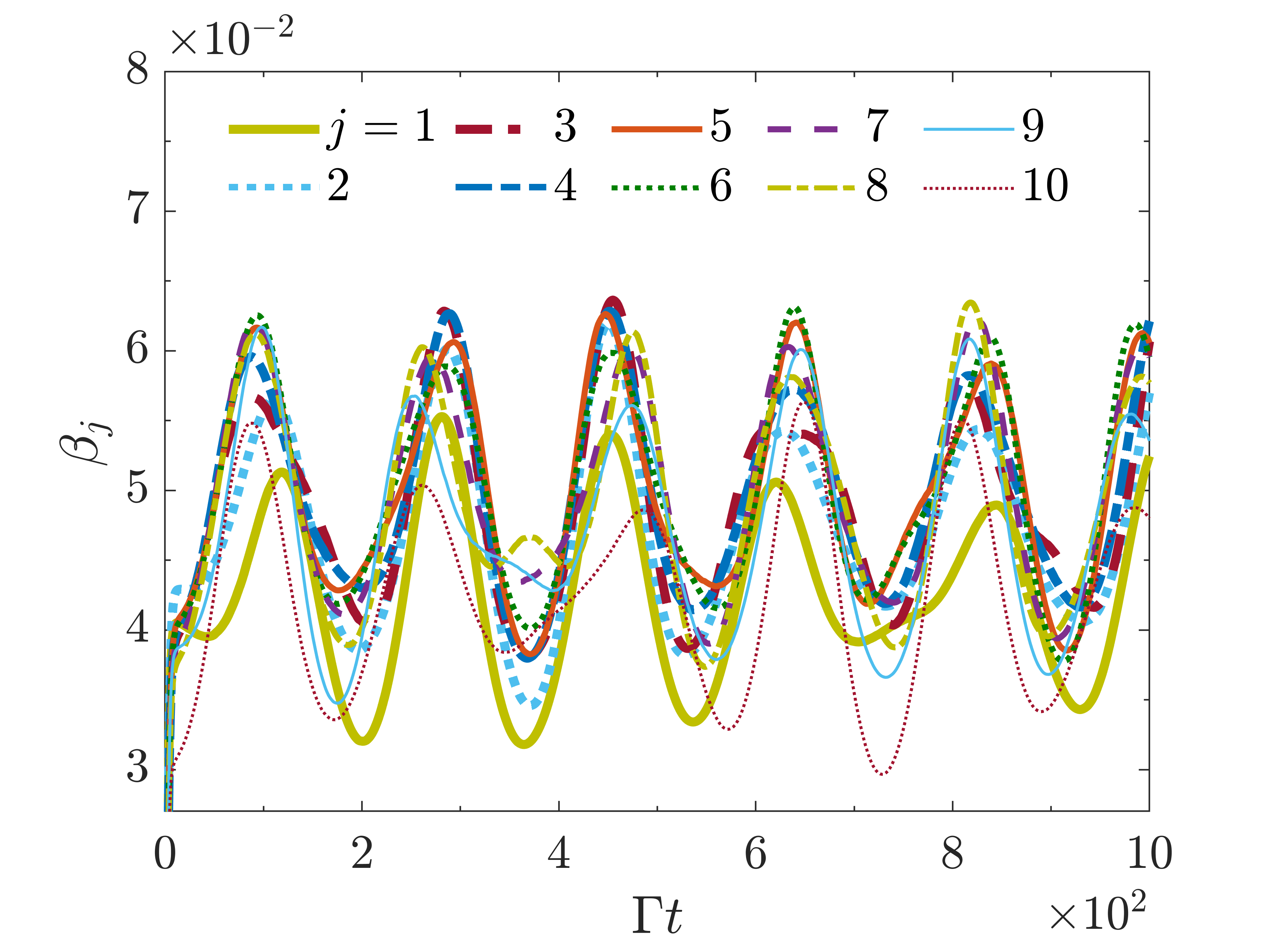}
\caption{Evolution of the dipole amplitude $\beta_j$ over time, for a chain of $N=10$ atoms. Simulation realized for a detuning $\Delta=-0.3\Gamma$ and a pump strength $\Omega=0.1\Gamma$. The atomic chain was here initialized with atoms separated by $\lambda$, with a tilt of $0.03\lambda$ toward positive $z$ of four atoms ($j=1,\ 3,\ 8,\ 10$): This breaking of symmetry allows to visualize the distinct dynamics of the $10$ dipoles.}
\label{fig:dipasync}
\end{figure}

A hint on the origin of this peculiar behaviour, as compared to a pair of atoms, can be found in the evolution of the atomic dipoles. Indeed a close analysis of the dynamics shows that for $N\geq 3$, differently than for $N=2$, the dipoles do not evolve synchronously, see Fig.~\ref{fig:dipasync}. One observes that the dipoles present substantial differences in their oscillations, both in terms of amplitude and oscillations maxima.

This lack of synchronization of the dipoles has strong consequences on the macroscopic dynamics, as revealed by comparing the full dynamics of Eqs \eqref{eq:betaj} and \eqref{eq:rj} to a synchronized ansatz, obtained by substituting in Eq.\eqref{eq:rj}  the values of the dipole amplitudes $\beta_j$ by their average $\overline{\beta}=(1/N)\sum_j\beta_j$.

\begin{figure}[!ht]
    \centering
    \includegraphics[width=\linewidth]{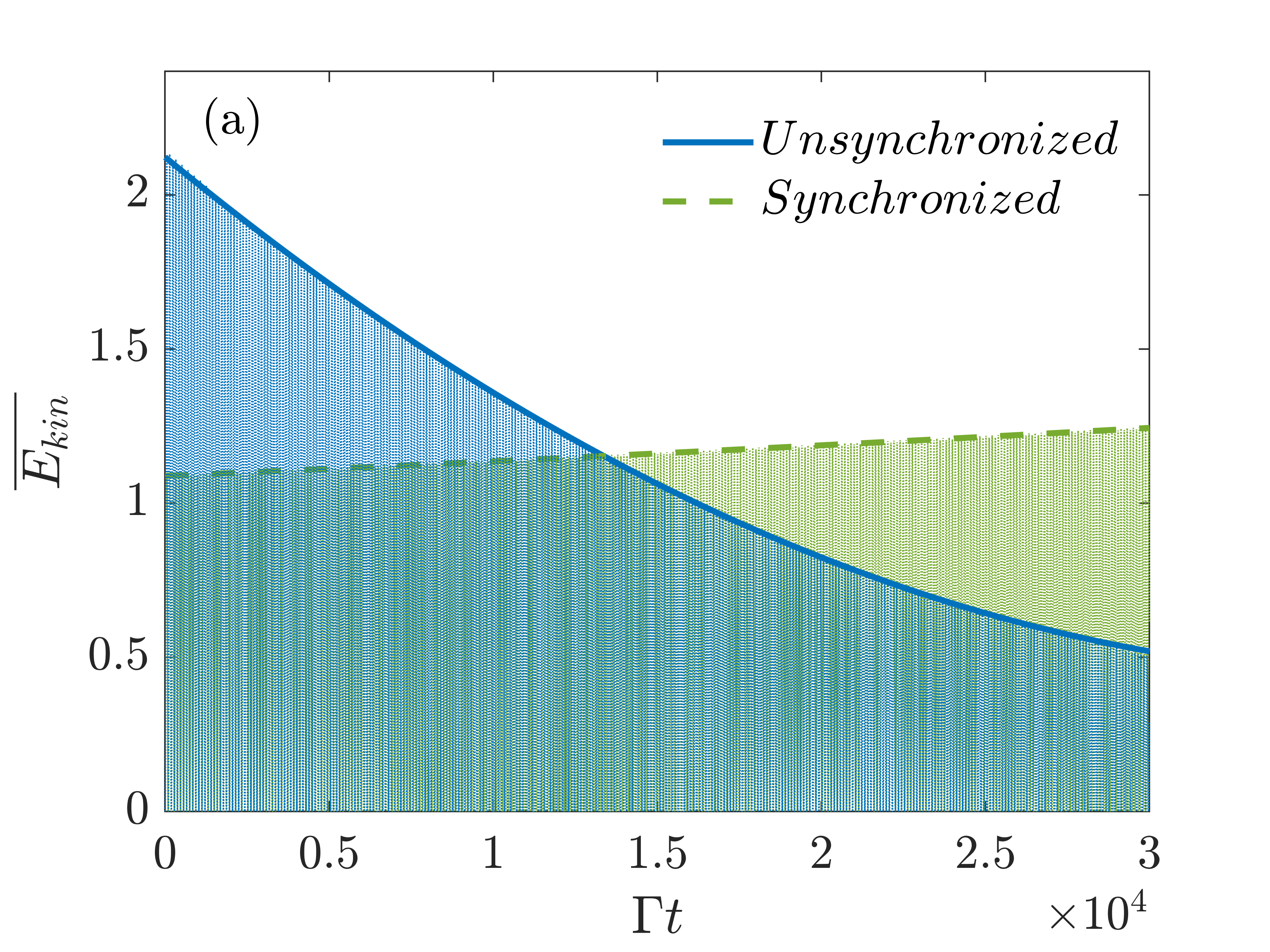}
    \includegraphics[width=\linewidth]{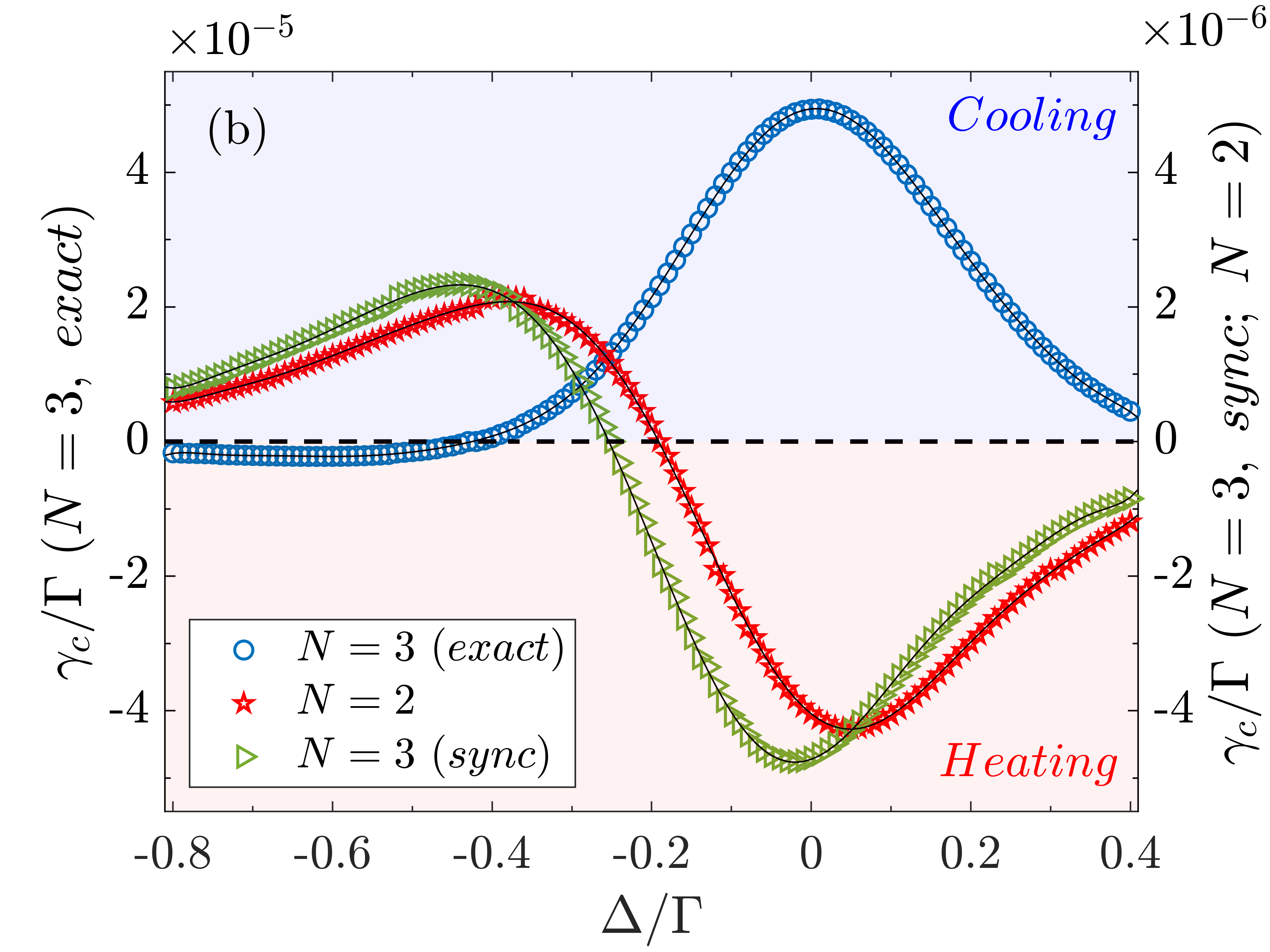}
    \caption{(a) Dynamics of the kinetic energy for $N=3$ atoms, comparing the full dipole dynamics of Eqs.(\ref{eq:betaj}) and (\ref{eq:rj}) with the synchronized obtained from the synchronization ansatz. Simulations realized with the detuning of optimal cooling for the exact case, $\Delta\approx 0.01\Gamma$. (b) Cooling/heating rate $\gamma_c/\Gamma$ for $N=3$ atoms, for the full dynamics ('exact') and imposing synchronized dipoles ('sync'), and for $N=2$ (the two dipoles spontaneously synchronize). The cooling/heating rate has been calculated using the evolution of the envelope of the kinetic energy until it reaches the $90/110\%$ of its initial value.}
    \label{fig:SyncvsUnsyn}
\end{figure}

As shown in Fig.~\ref{fig:SyncvsUnsyn}(a), close to resonance the synchronized dynamics presents a heating trend, whereas the full coupled dynamics exhibits a damping of the kinetic energy over time. The systematic comparison presented in Fig.~\ref{fig:SyncvsUnsyn}(b) confirms that using the synchronized ansatz, a chain of $N=3$ atoms displays the features of Doppler cooling (we checked that larger chains present a similar behaviour, up to a shift in the detuning that optimizes the cooling): Cooling is achieved only for negative detuning, and is maximal for $\Delta\approx-\Gamma/2$, whereas resonant light strongly heats the system. Differently, the $N\geq 3$ coupled dipole dynamics obtained from Eqs. (\ref{eq:betaj}-\ref{eq:rj}) exhibits a cooling which is maximum at resonance, but also significantly larger than for the synchronized case. Unfortunately, the asynchronous nature of this dynamics makes it very challenging to analyze it in more details, as one would need to deal with $N$ internal and $N$ external coupled degrees of freedom. Hence, despite the apparent complexity that the system dynamics presents, it is quite remarkable that this lack of synchronization results in a cooling rate much larger than the one encountered for synchronous dipoles.

\section{Impact of the spontaneous emission and velocity capture range\label{sec:fluct}}

Let us now discuss in more details the effect of heating due to spontaneous emission on the trapping. Considering that this process is dominated by the trapping beam $\Omega_0$, the rate of kinetic energy induced by spontaneous emission is:
\begin{equation}
\left(\dfrac{\delta 
E}{\delta t}\right)_{\mathrm{SE}}=\frac{\hbar\omega_r\Gamma}{3}\frac{\Omega_0^2}{\Gamma^2+4\Delta^2}.
\end{equation}
where the factor $1/3$ comes from the fact that the atomic recoil is distributed over the three spatial directions. Due to its oscillating nature (see Fig.\ref{fig:potential}(a)), the potential minimum is the opposite of its maximum (as the potential is here defined to be zero at large distances), so the heating has to overcome a barrier twice larger than the minimum potential, $\Delta U=2|U_{\mathrm{min}}|$, where $U_{\mathrm{min}}$ is provided by Eq.(\ref{Umin}).

\begin{figure}[h!]
\centering
\includegraphics[width=0.96\columnwidth,height=5.7cm]{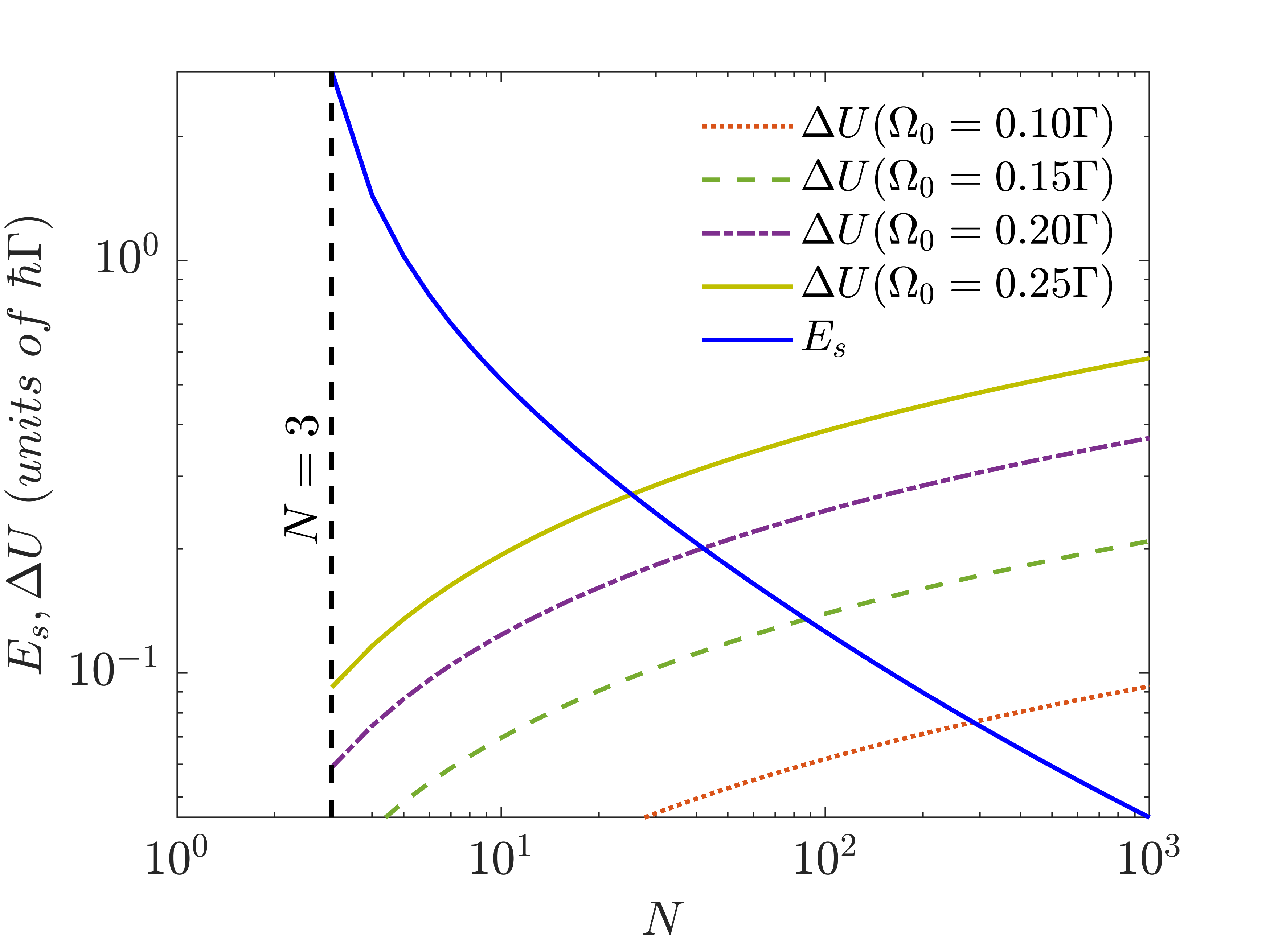}
\caption{Steady-state energy $E_s$ obtained from the Langevin equation and potential barrier $\Delta U$, as a function of the number of particles and for different pump strength $\Omega/\Gamma$.}
\label{fig:stab}
\end{figure}

Considering the exponential decay of the kinetic energy over time observed in the simulations, it is reasonable to include the cooperative cooling as a linear damping force, which leads to the following equation for the kinetic energy:
\begin{equation}
\frac{dE}{dt}=-\gamma_c E+\left(\dfrac{\delta E}{\delta t}\right)_{\mathrm{SE}}.
\end{equation}
where $\gamma_c$ is given by Eq.(\ref{gamma_c}). The steady-state energy is thus given by
\begin{eqnarray}
E_{\mathrm{s}}&=&\frac{1}{\gamma_c}\left(\dfrac{\delta E}{\delta t}\right)_{\mathrm{SE}}
\\ &\approx &\frac{0.83}{\ln N-0.8}\frac{\hbar\Gamma}{1+4(\Delta_c/\Gamma)^2}.
\end{eqnarray}
Stability is achieved when $E_{\mathrm{s}}<\Delta U$. For instance, for $\Omega_0/\Gamma=0.2$ stability should be reached for $N\geq 40$, see Fig.\ref{fig:stab}. While increasing pump strength suggests that a lower number of atoms is necessary to reach stability, it actually challenges the validity of the linear optics approximation~\cite{Williamson2020}, just as in the case of Doppler cooling.

In a similar way, we can estimate the velocity capture range $\Delta v$, assuming that the initial kinetic energy must be smaller than the potential barrier for the atoms to become trapped: $(m/2)(\Delta v)^2<2|U_{\mathrm{min}}|$. Using Eq.(\ref{Umin}), we obtain
\begin{equation}
    k\Delta v<2.4\sqrt{(\omega_r/\Gamma)\ln N}\,\Omega_0.
\end{equation}
For example, for \textsuperscript{87}Rb atoms on the D2 line ($5^2S_{1/2} \rightarrow 5^2P_{3/2}$ transition) one has that $\omega_r/\Gamma\sim 6\times 10^{-4}$, allowing to attain  $k\Delta v/\Gamma\sim 0.06\sqrt{\ln N}(\Omega_0/\Gamma)$. This can be compared to the values for Doppler cooling in optical molasses, $k\Delta v/\Gamma\sim 1$, and for Sisyphus cooling, $k\Delta v/\Gamma\sim\sqrt{\omega_r/\Delta_0}(\Omega_0/\Gamma)$~\cite{Dalibard1989}: The present cooperative cooling mechanism thus shares a scaling closer to Sisyphus cooling although, as discussed earlier, it is more efficient close to resonance. Finally, we point out that since the potential is self-generated by the atoms, this value is a simple estimation of the order of magnitude of the capture velocity, and a detailed study of the microscopic dynamics is necessary to obtain a precise value.

\section{Conclusions and Perspectives}

In conclusion, we have here shown that one-dimensional chains of cold atoms present a cooling mechanism which grows logarithmically with the system size. It relies on the presence of a resonance, and is thus absent from dielectric scatterers. Differently from other cooling mechanisms of cold atoms, the atomic chains are here most efficiently cooled very close to resonance, despite the strong spontaneous emission. A promising consequence is that chains of a few dozens of cold atoms could become stable thanks to this internal mechanism, without additional stabilizing mechanism such as optical molasses. Such self-organization mechanism could be probed, for example, using techniques inspired from Bragg scattering~\cite{vonCube2006}. 

This cooperative cooling mechanism, here discussed in the context of a purely one-dimensional system, may be even more promising for two-dimensional systems. Indeed, the $1/r$ term of the dipole-dipole interaction leads to a scaling as $\ln N$ in one dimension, but in two-dimensional systems the same argument will lead to a scaling as $\sqrt{N}$. Indeed, in a 2D lattice of atoms of edge $~\sqrt{N}$, the cooperativity parameter for an atom at site $(i,j)$  scales as $\sum_{l,m=1}^{\sqrt{N}} 1/\sqrt{(l-i)^2+(m-j)^2} \sim \sqrt{N}$. The OB forces are then expected to overcome in a more efficient way the fluctuations due to spontaneous emission, making two-dimensional self-generated lattices even more robust. One may also consider manipulating the balance between the OB potential and the spontaneous emission by taking advantage of the more complex internal structure of the atoms, using a Electromagnetically Induced Transparency configuration~\cite{Kocharovskaya1986,Morigi2000}. 

Finally, the self-cooling effect observed in our chain of atoms, connected by the exchanged photons scattered off the transverse driving fields, presents some analogies to collective cavity cooling in a high-finesse optical resonator~\cite{Ritsch2013}. These cavity self-organization effects have been suggested and studied theoretically by different groups in the 2000s~\cite{Domokos2002,Jager2016} and experimentally by Black et al.~\cite{Black2003} and, later, by Brennecke et al.~\cite{Brennecke2008}. Similarly, the atoms self-organize and cool into a self-generated potential with $\lambda$-spacing, built from the field scattered from a transverse drive. The main difference with free space scattering is that the cavity pre-selects a single mode of the electromagnetic field, in addition to strongly recycling the photons in some cases. 
More recently, it was suggested~\cite{Ostermann2016} that a single, strongly populated mode can spontaneously emerge also in free space from cooperative scattering by the atoms, which presents some analogies with the synchronization issue observed in the present work. An important difference of our work is that addressing all vacuum modes leads to a more complex system, for which the analogy with a single-mode approach remains to be demonstrated. Furthermore, we have observed that the adiabatic approximation, which could be used to simplify drastically the system by suppressing the fast timescale, has important consequences on the long-term stability of the atomic chain. In this context, the present work can be considered another step toward bridging the gap between free-space and cavity-based self-organization of cold atoms.

\section{Acknowledgements}

We thank Robin Kaiser and Carlos E. Maximo for fruitful discussions. This work was performed in the framework of the European Training Network ColOpt, which is funded by the European Union (EU) Horizon 2020 program under the Marie  Sklodowska-Curie action, grant agreement 721465. R.B. benefited from Grants from S\~ao Paulo Research Foundation (FAPESP) (Grants Nrs.~18/01447-2, 18/15554-5 and 19/12842-2) and from the National Council for Scientific and Technological Development (CNPq) Grant Nrs.~302981/2017-9 and 409946/2018-4.

\bibliography{references}

\end{document}